\documentclass[sn-nature]{sn-jnl}%

\usepackage{pifont}%
\usepackage{graphicx}
\usepackage[export]{adjustbox}\usepackage{multirow}%
\usepackage{amsmath,amssymb,amsfonts}%
\usepackage{amsthm}%
\usepackage{mathrsfs}%
\usepackage[title]{appendix}%
\usepackage{xcolor}%
\usepackage{textcomp}%
\usepackage{manyfoot}%
\usepackage{booktabs}%
\usepackage{algorithm}%
\usepackage{algorithmicx}%
\usepackage{algpseudocode}%
\usepackage{float}
\usepackage{siunitx}
\usepackage{amsmath}
\usepackage{hyperref}
\usepackage{listings}%
\usepackage{comment}
\usepackage{svg}
\usepackage{threeparttable}
\usepackage{colortbl}

\def\figureautorefname{Figure}

\begin{document}

\title[Omnidirectional gradient force optical trapping in dielectric nanocavities by inverse design]{Omnidirectional gradient force optical trapping in dielectric nanocavities by inverse design}

\author*[1,2]{\fnm{Beñat} \sur{Martinez de Aguirre Jokisch}}\email{bmdaj@dtu.dk}

\author[2,3]{\fnm{Benjamin} \sur{Falkenberg Gøtzsche}}\email{bfago@dtu.dk}

\author[2,3]{\fnm{Philip} \sur{Trøst Kristensen}}\email{ptkr@dtu.dk}

\author[2,3]{\fnm{Martijn} \sur{Wubs}}\email{mwubs@dtu.dk}

\author[1,2]{\fnm{Ole} \sur{Sigmund}}\email{olsi@dtu.dk}

\author[1,2]{\fnm{Rasmus} \sur{Ellebæk Christiansen}}\email{raelch@dtu.dk}

\affil*[1]{\orgdiv{Department of Civil and Mechanical Engineering}, \orgname{Technical University of Denmark}, \orgaddress{\street{Nils Koppels Allé Building 404}, \city{Kongens Lyngby}, \postcode{2800}, \country{Denmark}}}

\affil[2]{\orgdiv{NanoPhoton - Center for Nanophotonics}, \orgname{Technical University of Denmark}, \orgaddress{\street{Ørsteds Plads 345A}, \city{Kongens Lyngby}, \postcode{2800}, \country{Denmark}}}

\affil[3]{\orgdiv{Department of Electrical and Photonics Engineering}, \orgname{Technical University of Denmark}, \orgaddress{\street{Ørsteds Plads Building 343}, \city{Kongens Lyngby}, \postcode{2800}, \country{Denmark}}}


\abstract{Optical trapping enables precise control of individual particles of different sizes, such as atoms, molecules, or nanospheres. Optical tweezers provide free-space omnidirectional optical trapping of objects in laboratories around the world. As an alternative to standard macroscopic setups based on lenses, which are inherently bound by the diffraction limit, plasmonic and photonic nanostructures promise trapping by near-field optical effects on the extreme nanoscale. However, the practical design of lossless waveguide-coupled nanostructures capable of trapping sub-wavelength-sized particles in all spatial directions has until now proven insurmountable. In this work, we demonstrate an omnidirectional optical trap realized by inverse-designing fabrication-ready integrated dielectric nanocavities. The sub-wavelength optical trap is designed to rely solely on the gradient force and is thus particle-size agnostic. In particular, we show how a trapped particle with a radius of 15 nm experiences a force strong enough to overcome room-temperature thermal fluctuations. Furthermore, through the robust inverse design framework, we tailor manufacturable devices operating at short-wave infrared and near-infrared wavelengths. Our results open a new regime of levitated optical trapping by achieving a deep trapping potential capable of trapping single sub-wavelength particles in all directions using optical gradient forces. We anticipate potentially groundbreaking applications of the optimized optical trapping system for biomolecular analysis in aqueous environments, levitated cavity-optomechanics, and cold atom physics, constituting an important step towards realizing integrated bio-nanophotonics and mesoscopic quantum mechanical experiments.}


\maketitle
\section{Introduction}

Optical tweezers are versatile tools for interdisciplinary research due to their precise control and manipulation of micron-sized objects \cite{ashkin_acceleration_1970}. With applications ranging from studies in microbiology  
\cite{sudhakar_germanium_2021, ashkin_optical_1987}
to fundamental physics research \cite{gieseler_subkelvin_2012,dholakia_colloquium_2010,donato_light-induced_2016}, the optical tweezer is a decorated research tool. 
Undoubtedly, the success of the optical tweezer is owed to its ability to omnidirectionally trap objects by use of a single laser; however, since it is based on free-space optics, it is inherently bound by the diffraction limit. This limit can result in prohibitive power requirements to trap nanometer-scaled particles. As shown in \cite{marago_optical_2013}, a micrometer-sized polystyrene sphere can be stably trapped — stable against thermal diffusion of the particle — by an optical tweezer with a fraction of a milliwatt, while a 10 nm sphere requires 1.5 W. To overcome this power limitation, nanostructured metallic devices have been employed \cite{ mestres_unraveling_2016, jones_raman_2015, pang_optical_2012}, which can achieve stable trapping of nanometer-scaled particles at much lower powers, utilizing the strongly localized optical forces of plasmonic resonances. Such plasmonic devices have successfully been applied to the optical trapping of single proteins \cite{pang_optical_2012}, and single particle Raman spectroscopy \cite{jones_raman_2015}, among others. However, inherent losses in metals lead to heating and potential stability concerns of the electromagnetic resonator or adverse effects on the trapped objects \cite{xu_all-dielectric_2019, hernandez-sarria_toward_2021, mejia-salazar_plasmonic_2018, wang_plasmonic_2012}. To solve these issues, near-lossless nanostructured dielectric optical traps have been proposed \cite{xu_all-dielectric_2019, conteduca_fano_2023, hernandez-sarria_toward_2021, manka_simulation_2024}. Experimental realizations of dielectric traps include the trapping of single quantum dots \cite{xu_all-dielectric_2019} and the trapping of particles utilizing Fano resonances \cite{conteduca_fano_2023}, to name a few. \newline

Recent developments in the design and fabrication of dielectric bowtie-based nanocavities suggest access to hitherto unexplored field strength in dielectrics with deeply sub-wavelength light confinement \cite{hu_experimental_2018, albrechtsen_nanometer-scale_2022, babar_self-assembled_2023}. Bowtie nanocavities have previously been used as efficient optical traps \cite{xu_optical_2018, yoon_non-fluorescent_2018, brunetti_nanoscale_2022}, by utilizing the strong field enhancements at the material interfaces \cite{hu_design_2016, choi_self-similar_2017, albrechtsen_two_2022}. However, this field enhancement inevitably makes the particle stick to the resonator walls \cite{xu_optical_2018, brunetti_nanoscale_2022}, impeding omnidirectional trapping with gradient forces. For mesoscopic quantum mechanical experiments, any contact with the resonator material is detrimental to the coherence of the prepared state \cite{gieseler_subkelvin_2012}, and in biomolecular analysis, it may lead to undesired charge reconfiguration \cite{ishida_importance_2020}.\newline

In this work, we directly address the issue of omnidirectional trapping in nanostructures by inverse-designing dielectric nanocavities that trap particles with sizes significantly below the diffraction limit. A recent study \cite{manka_simulation_2024} addresses the issue of omnidirectionality for a limited range of particle sizes, by utilizing gradient forces and self-induced back-action (SIBA) effects \cite{neumeier_self-induced_2015}. Here, by only relying on gradient forces, we deterministically tailor the device geometry of a nanostructure to feature a particle-size independent omnidirectional trapping potential. 
The inverse design process relies on topology optimization (TO), a design optimization method widely used in the design of optical applications like waveguides \cite{jensen_systematic_2004, wang_robust_2011}, cavities \cite{wang_maximizing_2018, liang_formulation_2013, albrechtsen_nanometer-scale_2022}, demultiplexers \cite{piggott_inverse_2015}, microresonators \cite{ahn_photonic_2022}, and more. Experimental evidence \cite{albrechtsen_nanometer-scale_2022} shows that topology-optimized structures can directly incorporate manufacturing constraints \cite{zhou_minimum_2015,li_structural_2016}, ensuring precise fabrication of optimized device blueprints. 
In a recent work, a plasmonic optical trap was designed using TO by maximizing the electric-field strength at the center of a cavity \cite{nelson_inverse_2024}, yielding the well-known bowtie-like structure, that can trap particles at the material interfaces. In this work, we employ a novel TO scheme based on fitting the electromagnetic field profile to a desired Gaussian shape, enabling omnidirectional trapping within a specified volume. 
This scheme represents a versatile framework that can design optical traps for arbitrary material platforms, wavelengths, particle choices, and homogeneous background media. Using this framework we design two different waveguide-coupled and fabrication-ready devices at different wavelengths and homogeneous background materials, that may pave the way for future on-chip levitated optomechanics, on-chip biomolecular analysis, and cold-atom-based quantum many-body systems \cite{chang_colloquium_2018}.

\section{Inverse-designing omnidirectional trapping}\label{inverse_design}
To trap particles with a radius $R \ll \lambda$, where $\lambda$ is the wavelength of light, we may apply the dipole approximation \cite{novotny_principles_2012}. To show that our results apply to a wide range of lossless particles, we consider a non-resonant spherical particle with refractive index $n=2$ and radius $R=15$ nm, representative for both proteins ($n \simeq 1.6 $, $R \in [1$ nm$ - 100$ nm]) \cite{hernandez-sarria_toward_2021, erickson_size_2009} and semiconductor quantum dots ($n \simeq 2.4 $,  $R \in [2$ nm$ - 50$ nm]) \cite{hernandez-sarria_toward_2021,  wang_pseudopotential_1996, holewa_droplet_2022, protesescu_nanocrystals_2015}. 
For our choice of a lossless and non-resonant isotropic particle, scattering forces are negligible and the force may thus be well approximated by the gradient force. This conservative force can be described as the gradient of the trapping potential \cite{novotny_principles_2012}, 
\begin{equation}
    U(\mathbf{r})=-\frac{\alpha_{\text{R}}}{4}[\mathbf{E}^*(\mathbf{r})\cdot \mathbf{E}(\mathbf{r})]\,, 
\end{equation}
where $\mathbf{E}$ is the electric field phasor, $\mathbf{E}^*$ denotes its complex conjugate, and $\alpha_{\text{R}}$ is the real part of the spherical particle's polarizability, as given by the Clausius-Mossotti relation \cite{novotny_principles_2012}. Therefore, for a given particle in the nanostructure, the trapping potential is directly described by calculating the electric field distribution of the empty cavity. The dipole approximation may, however, break down for particles that are larger, exhibit loss, are resonant, or possess a higher refractive index, where additional effects like SIBA or scattering forces (e.g., radiation pressure or spin-curl forces) \cite{bustamante_optical_2021, novotny_principles_2012} may need to be accounted for. In post-analysis, we verify that our devices are unaffected by any of these effects.

To model optical trapping in the dipole approximation, we calculate the electric field of the empty cavity, by solving Maxwell's equations in the frequency domain, assuming time-harmonic behavior \cite{novotny_principles_2012}. The model is discretized and solved using the finite-element method with first-order Nedelec elements \cite{jin_finite_2014}. The trapping device consists of two silicon waveguides connected to a central design domain, in which TO is applied, see \hyperref[fig:1]{Figure 1a}. The central region contains a cylindrical air exclusion region with a radius of  $R_\text{exc}=300$ nm and thickness of 800 nm, in which the stable trap is located. One of the waveguides is excited with the fundamental mode at an input power of $P_\text{in}=60$ mW and at a free-space wavelength of $\lambda$. The wavelength is a freely selectable design parameter, determined by the desired application of the trap. Here as a demonstration, we inverse design a device in the short-wave infrared regime ($ \lambda = 1.55\,$\textmu m), and in one of the following sections, we inverse design an optical trap in the near-infrared regime ($\lambda=775$ nm). The material distribution in the design domain is optimized to obtain an electric field distribution that results in an omnidirectional trapping potential. Accordingly, we formulate our design problem as a continuous optimization problem where the material distribution is controlled by our design parameters and where we seek to minimize a Figure of Merit (FOM) that defines the difference of the electric-field norm with respect to a reference field $\mathbf{E}_{\text{ref}}$. 
This expression may be written as:
\begin{equation}\label{eq:FOM}
\text{FOM} \equiv \Phi= \sqrt{\int_{\Omega} \left[ \Theta\left(\frac{\left\|\mathbf{E}(\mathbf{r})\right\|}{\left\|\mathbf{E}(\mathbf{r}_0)\right\|}-\frac{\left\|\mathbf{E}_\text{ref}(\mathbf{r})\right\|}{\left\|\mathbf{E}_\text{ref}(\mathbf{r}_0)\right\|}\right)\right]^2 \mathrm{~d} \Omega}\,,
\end{equation}
where $\mathbf{r}_0$ is the center point in the design domain, $\Theta(x)$ is a smoothed Heaviside threshold function \cite{wang_projection_2011} and $\Omega$ is the optimization domain defined by the exclusion region, see \hyperref[fig:1]{Figure 1a}. The FOM takes the form of a $p$-norm with $p=2$, allowing a homogeneous penalization of the differences over the optimization domain $\Omega$. To ensure sub-wavelength omnidirectional trapping, we select the reference field to be a three-dimensional Gaussian potential with standard deviations $\sigma_{x}=\sigma_{y}=300$ nm and $\sigma_{z}=400$ nm, to ensure that it features a stable trapping minimum in all spatial directions. In this expression, the Heaviside projection $\Theta(x)$ ensures that the FOM promotes only electric field distributions as steep as, or steeper than the target field. If $\Theta(x)$ had not been applied then the difference to the target distribution would be minimized, not allowing for steeper trapping potentials and thus over-constraining the optimization problem by only fitting the target distribution. The standard deviations are chosen to match the dimensions of the cylindrical exclusion nanocavity with a volume $V_\text{cav}$ below the diffraction limit: $V_\text{cav}=0.22$ \textmu m$^3$$<$ $(\lambda/2)^3=0.465$ \textmu m$^3$. Additionally, we control the minimum depth of the trapping potential by adding a constraint to the optimization problem, which prescribes a minimum electric-field norm value at $\mathbf{r}_0$: $\log_{10}\left({||\mathbf{E}(\mathbf{r}_0)||}\right)\geq \gamma$, where $\gamma$ is a problem-dependent parameter defined in the Supporting Information (SI). By enforcing the constraint, the optimizer avoids local minima where the trapping potential has the correct shape but is not deep enough for stable trapping. To ensure the manufacturability of the device, we add a constraint to connect the design to the two waveguide ends \cite{li_structural_2016} as well as two minimum length scale constraints that act on the solid and void regions respectively \cite{zhou_minimum_2015}. To minimize the constrained FOM, we apply TO on the design domain, introducing one design variable per finite element in our discretized design domain, which is used to interpolate between air ($0$) and silicon ($1$). The design variables are fixed to only vary in the ($x,y$) plane and are linked in the $z$ direction \cite{christiansen_compact_2021}. We apply a filtering and thresholding procedure to regularize the design \cite{wang_projection_2011}. The filtered and thresholded design variables are related to the material's refractive index through a material interpolation scheme \cite{christiansen_non-linear_2019}. The optimized design is obtained by solving the topology optimization problem from a single uniform initial guess, using the globally convergent method of moving asymptotes as the optimizer \cite{svanberg_method_1987}. All inverse-design problems are solved using COMSOL Multiphysics \cite{COMSOL} executed on the DTU Computing Center HPC cluster \cite{DTU_DCC_resource}. For more information on the forward problem, the inverse design framework, and the full optimization problem see the SI.

\subsection{Inverse-designed nanocavity in the short-wave infrared}\label{sec:telecom}
\begin{figure}[ht!]
  \hspace*{-2cm} 
  \centering
  \includegraphics[width=160mm]{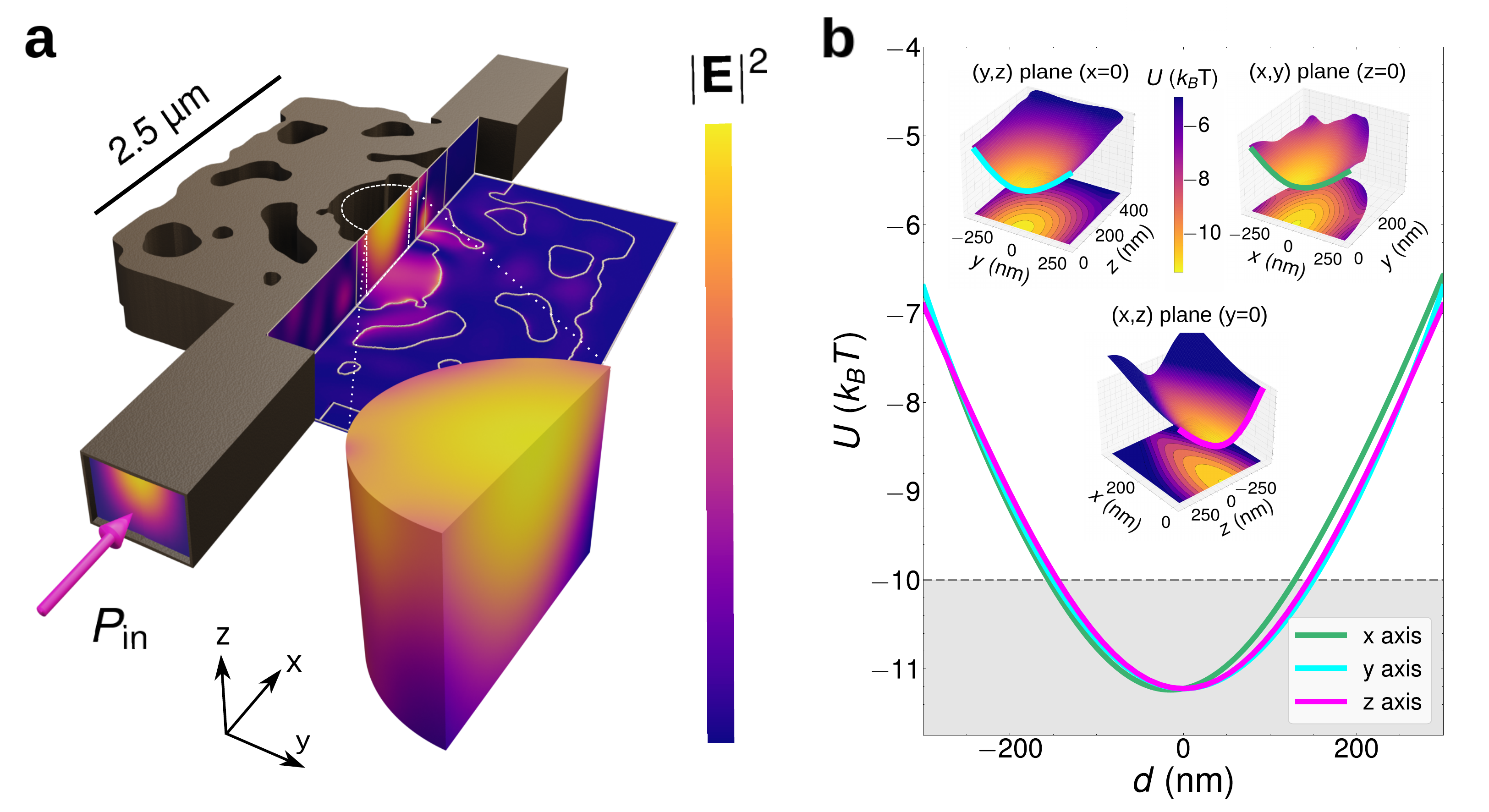}
  \caption{Optical response and trapping potential for the inverse-designed structure. \textbf{a} Rendering of the lower half of the optimized structure, with the electric-field intensity $|\mathbf{E}|^2$ response at $\lambda=1.55$ \textmu m, with a zoom-in in the optimization region ($\Omega$), when excited with the fundamental mode of the waveguide at an input power $P_\text{in}$ = 60 mW. \textbf{b} Trapping potential in the optimization region for the axial line- and plane-cuts as a function of the distance from the center ($d$). The stable trapping regime below $U = -10\, k_BT$ is shown in gray.}
  \label{fig:1}
\end{figure}

The inverse-designed omnidirectional trap and its key characteristics are presented in \autoref{fig:1}. In \hyperref[fig:1]{Figure 1.a} we show the electric-field intensity for the optimized structure in the short-wave infrared, which has a well-defined bell-shaped curve in all spatial directions. This is confirmed by observing the different plane- and line-cuts for the trapping potential in \hyperref[fig:1]{Figure 1.b}. One can identify a minimum of the potential at the optimization region's center, resulting in an omnidirectional trapping potential. Interestingly, the potential has a similar shape in all directions, meaning that the Heaviside projection $\Theta(x)$ in the FOM allowed the potential to become steeper than the reference Gaussian along the $z$-axis. For the 60 mW of optical input power and the reference particle with $R=15$ nm and $n=2$, the trapping minimum is stable against thermal fluctuations at room temperature ($T=300$ K), which conventionally requires a characteristic trapping depth of $U \simeq 10\, k_B T$ \cite{novotny_principles_2012}, where $k_B$ is the Boltzmann constant. Note that, as long as the dipole approximation is valid, and for high enough input power, the proposed device can stably trap arbitrarily small dielectric spheres in all directions. 

\begin{figure}[ht!]
  \hspace*{-1.0cm} 
  \centering
  \includegraphics[width=125mm]{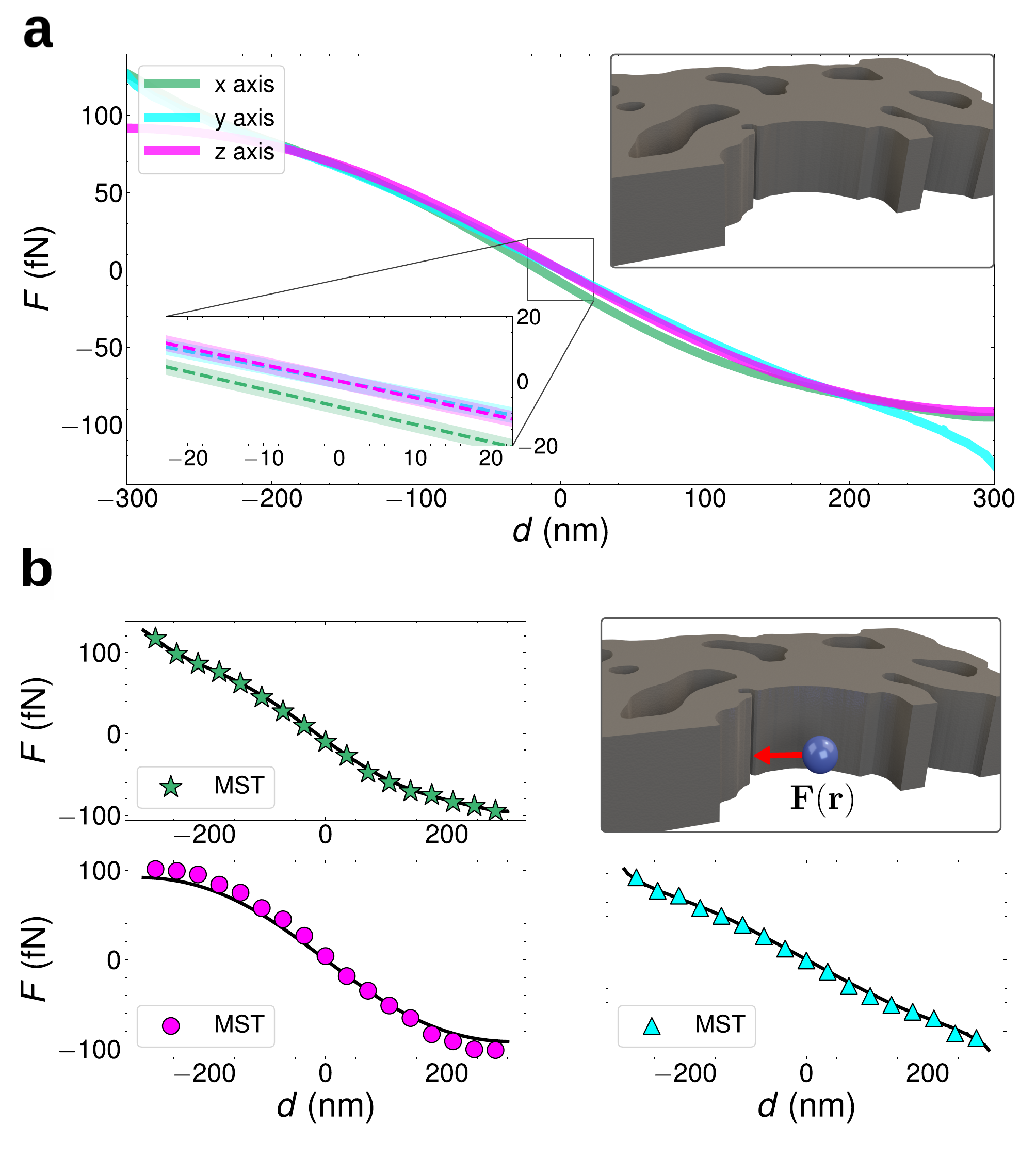}
  \caption{Trapping force calculation and Maxwell stress tensor (MST) validation. \textbf{a} Trapping force for the empty cavity in the dipole approximation for the axial components. \textbf{b} Force calculations via the MST for different particle positions, compared to the dipole approximation prediction (black line).}
  \label{fig:1_2}
\end{figure}

From the potential in the dipole approximation, we calculate the axial components of the force experienced by the particle in \hyperref[fig:1_2]{Figure 2.a}. Specifically, we show that close to the origin, the force becomes linear to a good approximation, and thus, the particle-trap system behaves as a linear spring-mass system. The force in the linear region can then be written as, $F_i(r_i)=\kappa_i\cdot r_i$, where $i \in \lbrace x,y,z \rbrace$ is an axis-index, $\kappa$ is a vector with the axial components of the trapping stiffness and $r_i$ is the position on the axis. Fitting the curves in the central region with a linear function yields trapping stiffnesses of $\kappa_x=0.53$ fN/nm, $\kappa_y=0.46$ fN/nm, and $\kappa_z=0.51$ fN/nm. It is possible to compare this the device to conventional optical tweezers, by comparing their trapping stiffnesses. The transverse trapping stiffness for a free-space optical tweezer is given by \cite{gieseler_subkelvin_2012}:
\begin{equation}
    \kappa_\text{OT}=4\pi^3\frac{\alpha_\text{R} P_\text{in}}{c \varepsilon_0}\frac{\text{NA}^4 }{\lambda^4}\,,
\end{equation}
 where the subscript OT refers to optical tweezers, $c$ is the speed of light, $\varepsilon_0$ is the vacuum permittivity, and NA is the numerical aperture. For a $\text{NA}=0.8$, as chosen {in \cite{gieseler_subkelvin_2012}}, the trapping stiffness is $\kappa_\text{OT}\simeq0.03$ fN/nm. In comparison to this reference our device outperforms conventional optical tweezers by more than an order of magnitude in all three directions.

To validate the dipole approximation, we solve the full Maxwell's equations with the sphere ($R=15$ nm and $n=2$) present in the optimized geometry, and calculate the force using the Maxwell Stress Tensor (MST) formalism \cite{novotny_principles_2012}. In \hyperref[fig:1_2]{Figure 2.b} we show the axial components of the force experienced by the particle when displaced in increments of $35$ nm in the three directions. By comparing the force derived from the MST calculation and the dipole approximation, we confirm that there is excellent agreement between the results and that there are no other dominant effects, such as SIBA \cite{neumeier_self-induced_2015,juan_self-induced_2009}, or scattering forces \cite{bustamante_optical_2021, novotny_principles_2012}.
By systematically increasing the particle size, it is shown \cite{jokisch_SPIE_2024} that the dipole approximation holds within a maximum relative error of $15\%$ for particles with sizes up to $R=100$ nm and refractive indices up to $n=8$, which shows the robustness of the presented methodology.

\subsection{Tuning the exclusion radius: from the omnidirectional trapping regime to the strong light confinement regime}\label{sec:excl_study}

\begin{figure}[h!]
  \hspace*{-1.5cm} 
  \centering
  \includegraphics[width=155mm]{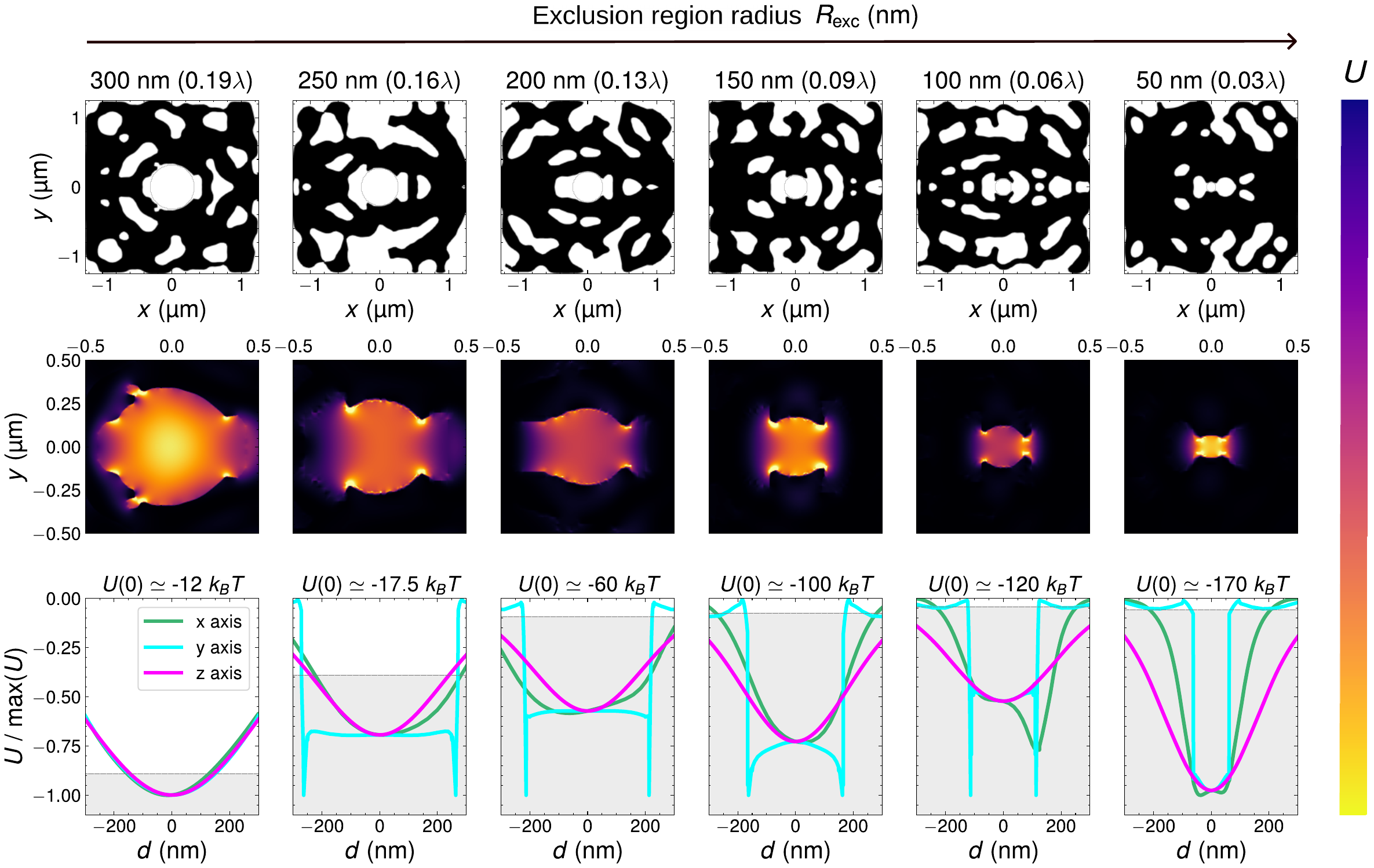}
  \caption{Projection of the nanocavity designs in the ($x,y)$ plane for decreasing radii of the exclusion region ($R_\text{exc}$), the spatial distribution of the optical trapping potential $U$ in the ($x,y)$ plane, and the projection of the max-normalized trapping potential on the cartesian axes. In the designs, black represents silicon, and white represents air. The stable trapping regime below $U=-10\, k_B T$ is shown in gray in the trapping potential projection onto the cartesian axes.}
  \label{fig:exc}
\end{figure}

One of the parameters that requires the most careful tuning in the inverse design procedure is the radius of the exclusion region ($R_\text{exc})$. In principle, to achieve the highest possible trapping stiffness, which is defined as the force exerted over a distance, one would try to make the exclusion region as small as possible. This would allow for stronger light confinement on a sub-wavelength scale, enhancing the trapping stiffness. However, through systematic optimizations selecting different exclusion radii, and accordingly different standard deviations of the target trapping potential ($\sigma_x$ and $\sigma_y$), we find a trade-off between the increased trapping stiffness and the possibility of achieving omnidirectional trapping with gradient forces. This is summarized in \autoref{fig:exc}, where for decreasing exclusion region radii and standard deviations of the target trapping potential, we optimize six nanocavity designs and calculate their trapping potential. The cavities optimized for smaller exclusion region radii can achieve strong field enhancements below the diffraction limit, providing deep trapping potentials. In fact, and based on the bowtie-like cavity design for the smallest exclusion region ($R_\text{exc}=50$ nm), we see that if dielectric material was allowed in the exclusion region, the optimizer would create extreme dielectric confinement bowtie-like cavities, similar to ones in \cite{albrechtsen_nanometer-scale_2022, babar_self-assembled_2023}. However, similar to bowtie-like cavity designs \cite{xu_optical_2018, yoon_non-fluorescent_2018, brunetti_nanoscale_2022}, when checking the trapping potential distribution for exclusion radii below $R_\text{exc}=200$ nm in \autoref{fig:exc}, it is clear that the strong field confinement located at the material interfaces does not allow for an omnidirectional trap in the center of the cavity, due to the spatial overlap of the field from the tip-like structures close to the center. This is similar to the double nanohole structure with a deeply sub-wavelength exclusion region found by another inverse design work \cite{nelson_inverse_2024} where the electric field was maximized at the center of the cavity, yielding a bowtie-like non-omnidirectional trapping potential. Interestingly, even for an exclusion radius of $R_\text{exc}=250$ nm the trapping potential in the $y$ direction starts to lose its concavity, which is further accentuated for smaller radii. In other words, there is a minimum length scale of the exclusion radius close to $R_\text{exc}\simeq 250$ nm, which sets a geometrical limit on omnidirectional trapping. By observing the projection of the trapping potential onto the cartesian axes it seems however, that one recovers stable trapping when going down to the smallest exclusion radius ($R_\text{exc}=50$ nm), but by looking at the two-dimensional profile of the trapping potential it is evident that a particle would be trapped at the nanocavity tips, where the field enhancement is strongest. In our work, we choose an exclusion radius $R_\text{exc}=300$ nm, since it simultaneously allows for an omnidirectional trapping without compromising the depth of the trapping potential or the trapping stiffness. This parameter choice allows to trap deeply sub-wavelength particles in a sub-wavelength exclusion region while overcoming thermal fluctuations.

\subsection{Vacuum surface forces: Casimir-Polder forces}

\begin{figure}[h!]
  \hspace*{-1.00cm} 
  \centering
  \includegraphics[scale=0.255]{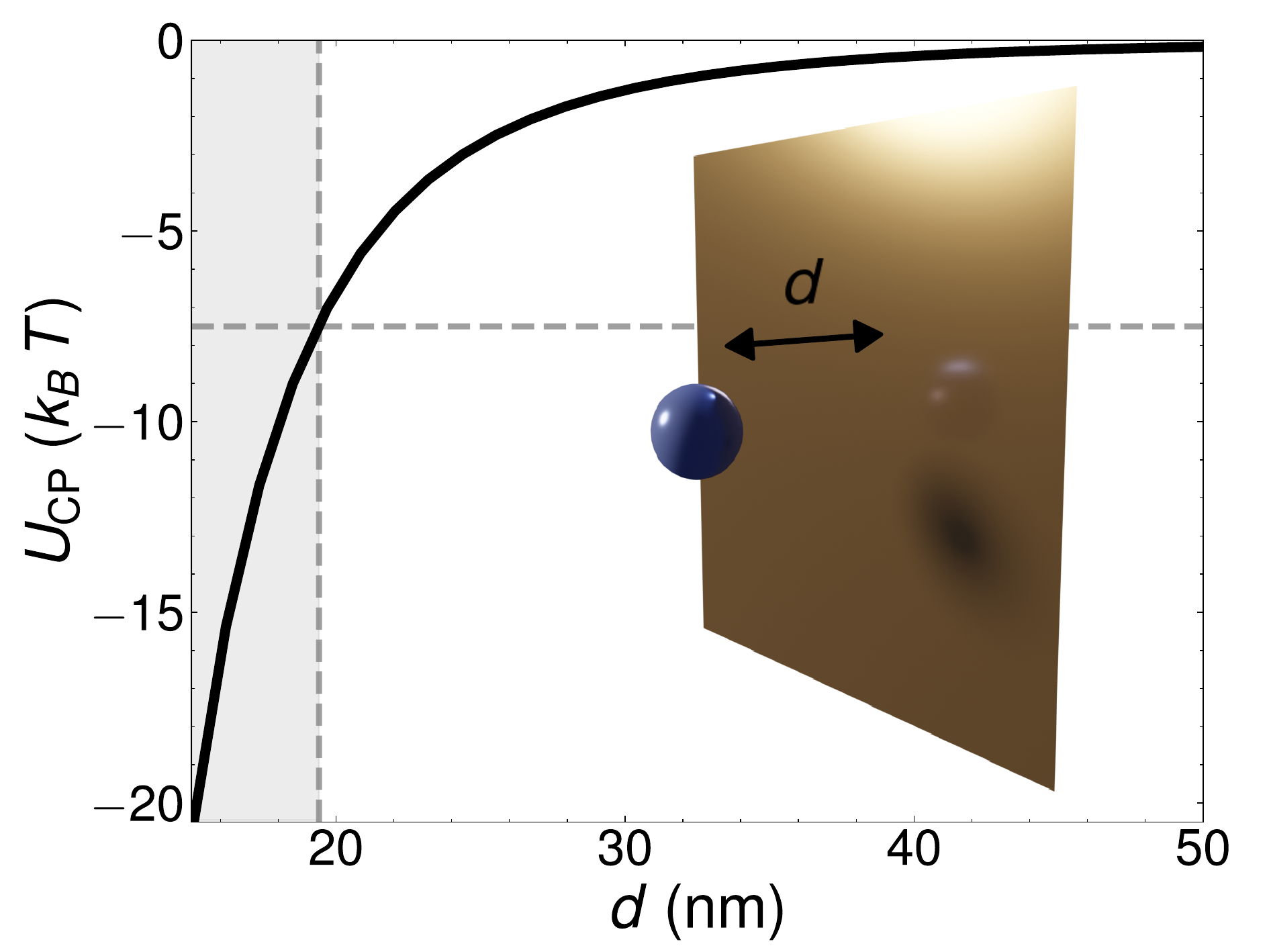}
  \caption{Casimir-Polder trapping potential ($U_\text{CP})$ as a function of the distance ($d$) between the center of a dielectric particle, with a refractive index $n=2$ and a radius of $R=15$ nm, and an infinite silicon plane with a refractive index $n=3.48$. Delimited by the dashed lines the region (in gray) where the CP potential starts to dominate the gradient force potential. Note that since the distance ($d$) is measured from the center of the particle we have excluded from the plot the region where $d<R=15$ nm.}
  \label{fig:3}
\end{figure}

We have shown that once the particle is at the center of the trap, it will be omnidirectionally trapped. However, loading the particle into the trap may potentially be jeopardized by other physical effects, such as the field enhancement observable at the bottom interface of the structure in \hyperref[fig:1]{Figure 1.a} or the Casimir-Polder (CP) forces near the material interfaces.  In section S3.1 of the SI we study the optical forces at the surfaces due to strong optical field enhancements, also known as lightning-rod effects \cite{albrechtsen_two_2022,choi_self-similar_2017}, and estimate the probability of a particle sticking to the surface due to these forces. In this section we study CP forces \cite{casimir_influence_1948}, which are vacuum forces that also exist in the absence of a light source, and result in a net attractive force to the surface that scales non-linearly with the distance between the particle and the surface. To calculate the probability of the particle getting stuck at the interface due to CP forces, we approximate the nanosphere as a polarizable point-dipole, close to an infinite dielectric plane, which is a good approximation since the CP forces are known to dominate only at short length scales in the order of 10 nm (see below). Given the high symmetry of this system, the CP energy landscape of this system reduces to a function of the distance $d$ between the
center of the particle and the plane. {Following \cite{Intravaia_2011}}, we calculate this function by integration of the Green tensor as formulated {in \cite{paulus_accurate_2000}}. We show the CP energy landscape as a normalized trapping potential in \autoref{fig:3} at room temperature (where $k_B\,T$ = $0.025$ eV), for a particle with a refractive index $n=2$ and a radius of $R=15$ nm, close to an infinite silicon interface with a refractive index $n=3.48$.

In the simplified model, the trapping potential diverges toward infinity as we approach the interface. To calculate the probability of the particle sticking to the interface due to the CP forces, we consider the stable and unstable trapping regimes in a cylinder extruded from the exclusion hole. Given that the potential at the edges of our exclusion domain is around  $U\simeq -7.5\, k_BT$, the CP potential will start to dominate the gradient force potential around $d\simeq 20$ nm from the interface. By assuming a simple positional averaging for the particle as it travels into the trap and then calculating the volume ratio between the volume where the particle feels an attractive force to the interface and the total volume, the probability of sticking to the surface is given by
\begin{equation}
    P_\text{stick}^\text{CP} \, [\%]= \frac{V_\text{CP}}{V_\text{total}} = \frac{ (R^2_\text{exc} - R^2_\text{CP} )}{  R^2_\text{exc} } \simeq 11.75 \% \,,
\end{equation}
where $V_\text{CP}$ is the volume fraction where the particle would stick to the interface due to the CP forces, $V_\text{total}$ is the total volume of the cylindrical exclusion, and $R_\text{CP}$ denotes the radius where the optical gradient forces dominate the CP forces. As shown in section S3.1 in the SI, compared to the lightning-rod effect, CP forces will more significantly affect the correct loading of the trap, as it will result in the particle sticking to the dielectric interface with a higher probability, given that $P_\text{stick}^\text{CP} \geq P_\text{stick}^\text{LR}$, where ``LR" denotes lightning-rod. It is important to note that the CP force is inherent to all air-mode cavities, and thus, it is not possible to avoid this source of particles sticking to the surface, only to minimize it by reducing the surface-to-volume ratio near the trapping region. For instance, if one reduced the exclusion region radius, then the surface-to-volume ratio would increase, resulting in even more sticking events due to the CP forces. For example, from the results in \autoref{fig:3}, an exclusion region radius of 20 nm yields a sticking probability of $P_\text{stick}^\text{CP}\simeq 90 \%$, which will approach unity as the exclusion becomes even smaller. Therefore, for deeply sub-wavelength design features, such as the ones in other dielectric nanocavities \cite{albrechtsen_nanometer-scale_2022, babar_self-assembled_2023}, the CP forces would entirely dominate the optical trapping, making levitated omnidirectional optical trapping unattainable. In other words, vacuum forces set a minimum length scale on the exclusion region radius for levitated omnidirectional trapping in nanostructures, which for this device is around $R_\text{exc}\simeq 20$ nm. For omnidirectionally trapping nanostructures, the minimum length scale set by CP forces can be reduced by increasing the input power, so that optical forces dominate vacuum forces. Therefore, this largely favors dielectric designs compared to
plasmonics, since in the considered wavelength range metals have a much larger imaginary
component of the relative permittivity, leading to significantly higher optical losses.

\subsection{Benchmarking trapping performance: normalized trapping stiffness}

In the optical trapping community, different metrics are used to benchmark optical trapping platforms of lossless particles; among others, power normalized trapping depth  \cite{manka_simulation_2024, nelson_inverse_2024, hernandez-sarria_numerical_2023, hernandez-sarria_toward_2021}, trapping force \cite{nelson_inverse_2024,hernandez-sarria_numerical_2023,xu_all-dielectric_2019, hernandez-sarria_toward_2021}, and trapping stiffness \cite{juan_self-induced_2009, mestres_unraveling_2016, bouloumis_enabling_2023, conteduca_fano_2023, gieseler_subkelvin_2012} are commonly used. These measures may depend on the physical properties of the trapped particle and its environment, the device footprint, or the input power, making comparisons difficult. Therefore, we propose a normalized trapping stiffness metric ($\eta_i$) for optical trapping that normalizes trapping stiffness to particle volume, particle-background permittivity contrast and input power,
\begin{equation}\label{eq:metric}
    \eta_i = \frac{\kappa_i \, \varepsilon_0}{\alpha_{\text{R}} \, P_\text{in}}\,,
\end{equation}
where $i \in \lbrace x,y,z \rbrace$ is an axis-index. This metric, which characterizes the power efficiency of optical forces per distance, enables one-to-one comparisons of optical trapping among different platforms. For our trap we find $\eta_x = 0.42 $ pN/(\textmu m$^4$$\cdot$\textmu W), $\eta_y = 0.36 $ pN/(\textmu m$^4$$\cdot$\textmu W) and $\eta_z = 0.40 $ pN/(\textmu m$^4$$\cdot$ \textmu W) along the three axial directions. Note that this metric is not applicable for trapping platforms with no light source, such as vacuum force trapping structures, since there is no input power ($P_\text{in}$) and the normalized trapping stiffness $\eta_i$ diverges. The metric could be modified for these devices by redefining \autoref{eq:metric} without including the input power.

To see how its trapping characteristics compare to other state-of-the-art trapping devices, we have gathered and compared the main trap parameters across several platforms in \autoref{tab1}, including optical tweezer, plasmonic, dielectric platforms, which may or may not rely on SIBA effects. Note that the devices relying on SIBA effects, are only expected to work for sufficiently large (and fixed) particles, unlike optical tweezers and this work, which can be used for all particle sizes as long as the dipole approximation is valid, enabling the trapping of deeply sub-wavelength particles. Therefore, this largely favors dielectric designs compared to
plasmonics, since in the considered wavelength range metals have a much larger imaginary
component of the relative permittivity, leading to significantly higher optical losses. Additionally, and similar to other dielectric devices \cite{manka_simulation_2024, mandal, Tonin}, we propose an integrated photonic device, meaning that it is waveguide-coupled, in contrast to many other devices that are excited out-of-plane by the use of lasers \cite{juan_self-induced_2009, bouloumis_enabling_2023, mestres_unraveling_2016, nelson_inverse_2023,hernandez-sarria_numerical_2023, xu_optical_2018, conteduca_fano_2023}. For these devices, where the laser excitation intensity is reported, we use the laser spot area to compute the total power in \autoref{tab1}.  Using the normalized trapping stiffness as indicator, we observe that devices relying upon plasmonic resonances can achieve stronger normalized trapping stiffnesses \cite{juan_self-induced_2009, bouloumis_enabling_2023, mestres_unraveling_2016, nelson_inverse_2023} than our inverse-designed device. However, plasmonic devices lack omnidirectional trapping, which leads to particles sticking to the surface, and suffer from heating problems due to the optical losses in metals, which can be detrimental in many trapping settings \cite{ishida_importance_2020, xu_all-dielectric_2019, hernandez-sarria_toward_2021, mejia-salazar_plasmonic_2018, wang_plasmonic_2012}. We have highlighted this problem in \autoref{tab1} by coloring the platforms that lose energy to heating in \textcolor{red}{red} and the ones that do not, in \textcolor{blue}{blue}. Regarding dielectric platforms \cite{Tonin, mandal, xu_optical_2018, conteduca_fano_2023}, we see that our platform offers normalized trapping stiffnesses in the same order of magnitude, while being the only non-SIBA solution offering optical trapping in all spatial directions. As previously discussed, for optical trapping nanostructures that do not utilize SIBA effects, there seems to be a trade-off between achieving stability in the spatial directions on the one hand, and the normalized trapping stiffness achievable by the optical platforms on the other hand. Compared to optical tweezers, which provide omnidirectional trapping, we achieve similar trapping stiffnesses at significantly lower input power, thus yielding higher normalized trapping stiffnesses. This demonstrates the usefulness of miniaturizing the free-space optics by means of a waveguide-coupled nanostructured device relying on near-field optical effects. This is also exemplified by comparing the normalized trapping stiffness $\eta_i$ in \autoref{tab1} for our vacuum device ($\lambda=1.55$ \textmu m) and aqueous device ($\lambda=775$ nm), which for the same particle size and input power, reveal more than an order of magnitude increase of $\eta_i$ by operating at smaller scales. 
\begin{table}[htbp]
\makebox[\linewidth][l]{%
        \hspace*{-1.75cm} 
\footnotesize
\begin{tabular}{@{}lllllllll@{}}
\toprule
	 Reference & $D$ &$\eta_i$ [pN/(\textmu m$^4$$\cdot$\textmu W)] & $\kappa$[fN/nm]  & $R$ [nm] & $n_\text{p}$ ($\omega$) & $n_\text{back}$ ($\omega$) & $P_\text{in}$ [mW] & Platform   \\
  \midrule

    \rowcolor[HTML]{FFDAB9}\cite{juan_self-induced_2009} & \textcolor[HTML]{8B4513}{1} &
  \textcolor{black}{1.95$\cdot10^4$ - 3.62$\cdot10^5$} & \textcolor{black}{6000 - 7000}& \textcolor{black}{25 - 50} & \textcolor{black}{1.58}& \textcolor{black}{1.33} &  \textcolor{black}{0.7 - 1.9}  &   \textcolor{red}{Plasmonic (SIBA)} \\


    \rowcolor[HTML]{FFDAB9}\cite{mandal}& \textcolor{black}{1} & \textcolor{black}{16.79 - 43.39} & \textcolor{black}{3300 - 8530} & \textcolor{black}{50} &  \textcolor{black}{1.59} & \textcolor{black}{1.33}  & \textcolor{black}{1000} &   \textcolor{blue}{Dielectric}\\

  \rowcolor[HTML]{FFDAB9}\cite{Tonin}& \textcolor{black}{1} & \textcolor{black}{13.18} & \textcolor{black}{300} & \textcolor{black}{250} &  \textcolor{black}{1.57} & \textcolor{black}{1.33}  & \textcolor{black}{1} &   \textcolor{blue}{Dielectric (SIBA)}\\
  
    \rowcolor[HTML]{FFFDD0}\cite{bouloumis_enabling_2023}& \textcolor[HTML]{FF8C00}{2} & \textcolor{black}{31.64} & \textcolor{black}{4.18} & \textcolor{black}{10} & \textcolor{black}{0.19 + 5.93i} & \textcolor{black}{1.33} &  \textcolor{black}{$9^*$} &    \textcolor{red}{Plasmonic (SIBA)} \\
    \rowcolor[HTML]{FFFDD0}\cite{mestres_unraveling_2016}& \textcolor[HTML]{FF8C00}{2} & \textcolor{black}{3.33} & \textcolor{black}{2.4}  & \textcolor{black}{30}  & \textcolor{black}{0.26 + 6.97i} & \textcolor{black}{1.33} &  \textcolor{black}{$1.9$}  &    \textcolor{red}{Plasmonic (SIBA)} \\
    \rowcolor[HTML]{FFFDD0}\cite{nelson_inverse_2023}& \textcolor[HTML]{FF8C00}{2} & \textcolor{black}{-} & \textcolor{black}{-}  & \textcolor{black}{14.5} & \textcolor{black}{1.6} & \textcolor{black}{1.33} & \textcolor{black}{105$^*$} &   \textcolor{red}{Plasmonic}\\
    \rowcolor[HTML]{FFFDD0}\cite{hernandez-sarria_toward_2021}& \textcolor[HTML]{FF8C00}{2} & \textcolor{black}{-} & \textcolor{black}{-} & \textcolor{black}{10 - 18}  & \textcolor{black}{2} & \textcolor{black}{1.33}&  \textcolor{black}{$100$}  &    \textcolor{blue}{Dielectric} \\
    \rowcolor[HTML]{FFFDD0}\cite{wang_-chip_2022}& \textcolor[HTML]{FF8C00}{2} & \textcolor{black}{-} & \textcolor{black}{-} &  \textcolor{black}{15 - 60} & \textcolor{black}{1.59} & \textcolor{black}{1} & \textcolor{black}{0.02}   &   \textcolor{blue}{Dielectric}\\
    
    \rowcolor[HTML]{FFFDD0}\cite{xu_optical_2018}& \textcolor[HTML]{FF8C00}{2} & \textcolor{black}{10.71 - 21.41} & \textcolor{black}{0.04 - 0.08} & \textcolor{black}{10} & \textcolor{black}{1.57} & \textcolor{black}{1.33} & \textcolor{black}{2.5$^*$} &   \textcolor{blue}{Dielectric}\\
    
    \rowcolor[HTML]{FFFDD0}\cite{conteduca_fano_2023}& \textcolor[HTML]{FF8C00}{2} & \textcolor{black}{0.06} & \textcolor{black}{1.19} & \textcolor{black}{50 - 75} &  \textcolor{black}{1.58} & \textcolor{black}{1.33}  & \textcolor{black}{105$^*$} &   \textcolor{blue}{Dielectric}\\

    \rowcolor[HTML]{FFFDD0}\cite{Tonin_1}& \textcolor{black}{2} & \textcolor{black}{-} & \textcolor{black}{-} & \textcolor{black}{250} &  \textcolor{black}{1.57} & \textcolor{black}{1.33}  & \textcolor{black}{100} &   \textcolor{blue}{Dielectric}\\



   \rowcolor[HTML]{CCFFCC} \cite{manka_simulation_2024}& \textcolor[HTML]{006400}{3} & \textcolor{black}{-} & \textcolor{black}{-} & \textcolor{black}{50} & \textcolor{black}{1.44} & \textcolor{black}{1} & \textcolor{black}{1} & \textcolor{blue}{Dielectric (SIBA)} \\
    \rowcolor[HTML]{CCFFCC}\cite{gieseler_subkelvin_2012}& \textcolor[HTML]{006400}{3} & \textcolor{black}{0.001 - 0.02} & \textcolor{black}{0.14 - 2.24$^{**}$} & \textcolor{black}{70} & \textcolor{black}{1.44} & \textcolor{black}{1} & \textcolor{black}{100} &  \textcolor{blue}{Optical tweezer} \\
    
    \midrule
    \rowcolor[HTML]{CCFFCC}\textbf{This work}&&&&&&&& \\
     \rowcolor[HTML]{CCFFCC} \textbf{$\lambda=$ 1.55 \textmu m} & \textbf{\textcolor[HTML]{006400}{3}} & \textbf{\textcolor{black}{0.36 - 0.42}} & \textbf{\textcolor{black}{0.46 - 0.53}} & \textbf{\textcolor{black}{15}}& \textbf{\textcolor{black}{2}} & \textbf{\textcolor{black}{1}} &\textbf{\textcolor{black}{60}}&  \textbf{\textcolor{blue}{Dielectric}} \\
    \rowcolor[HTML]{CCFFCC} \textbf{$\lambda=$ 775 nm}  & \textbf{\textcolor[HTML]{006400}{3}} & \textbf{\textcolor{black}{5.96 - 9.95}} & \textbf{\textcolor{black}{4.49 - 7.49}} & \textbf{\textcolor{black}{15}}& \textbf{\textcolor{black}{2}} & \textbf{\textcolor{black}{1.33}} &\textbf{\textcolor{black}{60}}&  \textbf{\textcolor{blue}{Dielectric}}
	\end{tabular}
 }
 \vspace*{4mm}
 \caption{Comparison of different trapping platforms 
 in terms of the number of stable trapping directions ($D$), normalized trapping stiffness ($\eta_i$), trapping stiffness ($\kappa$), sphere radius ($R$), frequency-dependent refractive index of the particle ($n_\text{p}$), frequency-dependent background refractive index ($n_\text{back}$), input power ($P_\text{in}$) and type of platform (plasmonics, dielectrics or optical tweezers). To emphasize the number of stable directions the rows have been colored accordingly, \colorbox[HTML]{FFDAB9}{{\textcolor[HTML]{8B4513}{orange}}} for 1 stable direction, \colorbox[HTML]{FFFDD0}{{\textcolor[HTML]{FF8C00}{yellow}}} for 2 stable directions and \colorbox[HTML]{CCFFCC}{{\textcolor[HTML]{006400}{green}}} for omnidirectional trapping with 3 stable directions. We also color the platform text to highlight which platforms lose energy to heating (in \textcolor{red}{red}) and which are nearly lossless (in \textcolor{blue}{blue}). The empty values denoted by "-" were not reported in the references, while we calculated the values marked with * using the light intensity and laser spot size. The value marked with ** was calculated using the fundamental mechanical frequencies of the particle's oscillation in the optical tweezer \cite{gieseler_subkelvin_2012} and its mass, which was calculated from the volume of the particle and the density of fused silica ($\rho=2.2$ g/cm$^3$).}. \label{tab1}%
\end{table}

\subsection{Inverse-designed nanocavities for omnidirectional trapping applications}\label{sec:app}

\begin{figure}[ht!]
  \hspace*{-1.5cm} 
  \centering
  \includegraphics[width=165mm]{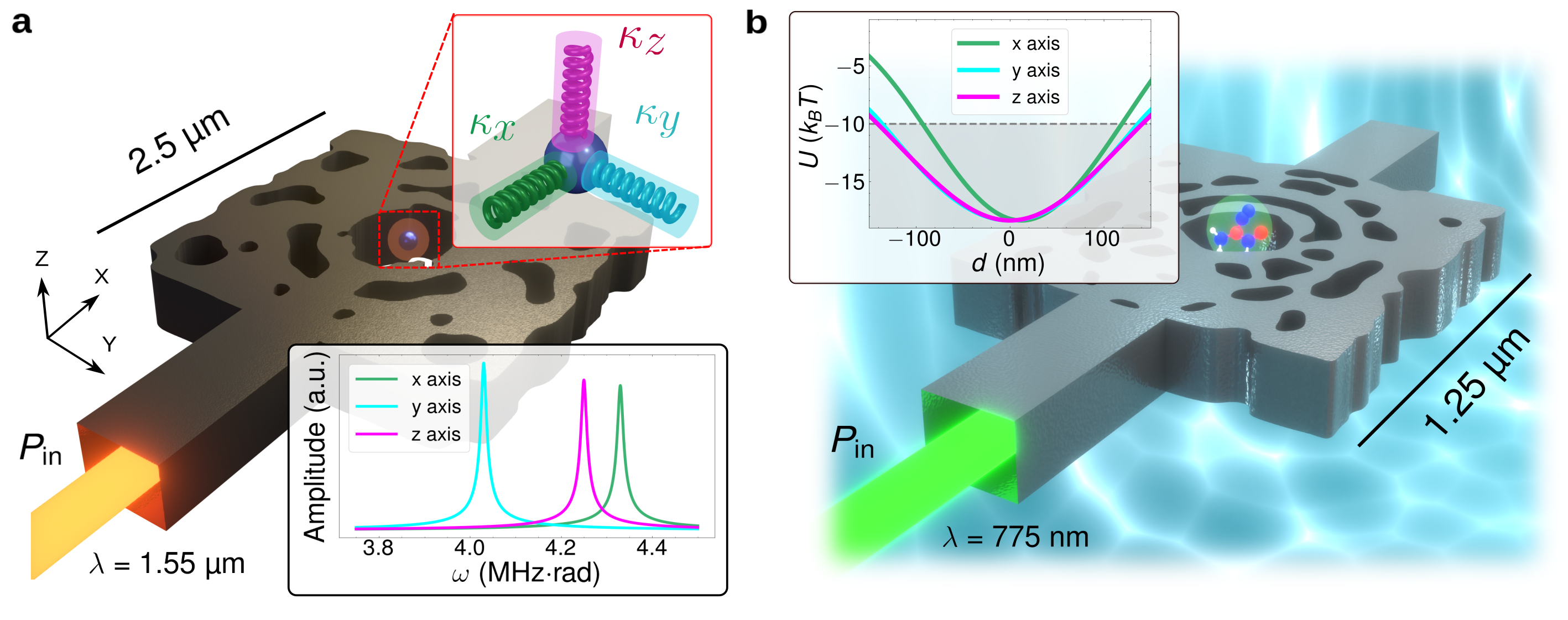}
  \caption{Rendering of the lower half of the inverse-designed structures for two excitation wavelengths. \textbf{a} Levitated cavity optomechanical setup, with the frequency response of the mechanical modes of the trapped particle in the three spatial axes. We assume a spectral broadening of $1.5$ kHz. \textbf{b} Integrated biophotonic setup in an aqueous environment, where a molecule (modeled as a particle with $n=2$ and $R=15$ nm) is trapped in a stable omnidirectional potential.}
  \label{fig:2}
\end{figure}

The miniaturized integrated circuit shown in \hyperref[fig:1]{Figure 1.a} bridges the omnidirectional trapping of optical tweezers with the near-field optics of ultra-compact nanostructures, thereby enabling a range of chip-scale applications. As an example, we demonstrate its potential in levitated cavity optomechanics. Owing to the harmonic potential close to the cavity center, we calculate the natural frequencies of the harmonic oscillator for the different axes as $\Omega_{0,i} = \sqrt{\kappa_i/m}$, where $m$ is the mass of the particle and $i \in \lbrace x,y,z \rbrace$ is an axis-index. Using a density of $\rho=2$ g/cm$^3$, which represents particles like proteins ($\rho \in [1.4 $ g/cm$^3$ - 1.5  g/cm$^3$]) \cite{fischer_average_2004} and quantum dots (e.g. $\rho_\text{Si}=2.33$ g/cm$^3$, $\rho_\text{GaAs}=5.32$ g/cm$^3$), we find $\Omega_{0,x}=4.33$ rad$\cdot$MHz, $\Omega_{0,y}=4.03$ rad$\cdot$MHz and $\Omega_{0,z}=4.25$ rad$\cdot$MHz. In the quantum-mechanical limit, the mean thermal occupancy of the mechanical energy states is given by $\langle n\rangle=k_B T / \hbar \Omega_0$ \cite{gieseler_subkelvin_2012}, where $\hbar$ is the reduced Planck constant. Resolving the quantum ground state requires $\langle n \rangle < 1$, which with the natural frequencies of this system, requires center-of-mass equilibrium temperatures of $T_x = 0.21$ mK, $T_y = 0.19$ mK and $T_z= 0.20$ mK. These temperatures are two orders of magnitude higher than for conventional optical tweezers \cite{gieseler_subkelvin_2012}, which combined with the particle-size agnostic omnidirectional trapping, facilitates mesoscopic quantum optomechanical experiments with optically trapped and cooled nanoparticles. One could potentially reach these temperatures by lowering the ambient temperature while employing methods like parametric feedback cooling in a vacuum chamber \cite{gieseler_subkelvin_2012}. A representation of the optomechanical system is shown in \hyperref[fig:2]{Figure 5.a}.

To demonstrate the versatility of the design framework, we move towards near-infrared frequencies and design a device suited for integrated biological sensing. For biological characterization in aqueous environments, it is crucial to operate within the biological window \cite{feng_perfecting_2021},  a spectral range where light has a long penetration depth in tissue, and water has low optical losses. This helps alleviate losing too much energy to heating, which could otherwise compromise the trapping setup  \cite{xu_all-dielectric_2019, hernandez-sarria_toward_2021, mejia-salazar_plasmonic_2018, wang_plasmonic_2012}. Although, the optical losses of water may be lower for wavelengths below the biological window, optical losses in silicon become larger, which would lead to a reduction in the efficiency of our device. To this end, we scale all geometric design parameters and the wavelength by half, shifting towards the near-infrared frequencies at $\lambda=775$ nm. Accordingly, we set the background refractive index to water and adjust the refractive index of silicon. With these new design parameters, we resolve the optimization problem and obtain the trap design in \hyperref[fig:2]{Figure 5.b}. 
This trap yields an omnidirectional stable trap for a single particle at the center of the design, with trapping stiffnesses of $\kappa_x=7.49$ fN/nm, $\kappa_y=4.49$ fN/nm, $\kappa_z=4.77$ fN/nm, for a particle with radius $R=15$ nm and for an input power of $P_\text{in}=60$ mW.  The normalized trapping stiffnesses for this device are $\eta_x = 9.95 $ pN/(\textmu m$^4$$\cdot$\textmu W), $\eta_y = 5.96 $ pN/(\textmu m$^4$$\cdot$\textmu W) and $\eta_z = 6.33 $ pN/(\textmu m$^4$$\cdot$\textmu W). These values are larger than those of the device operating in the short-wave infrared due to the smaller footprint of the device: given that both devices are fed with the same input power and trap exactly the same particle, halving all the geometric design parameters results in an increased trapping stiffness, since the stiffness is given as a force per unit distance. The inverse-designed optical trapping device is thus an integrated optical component capable of stably trapping biomolecules in all spatial dimensions in an aqueous solution. Note that omnidirectional trapping still holds for lower values of the refractive index of molecules (e.g. $n\sim1.5$) \cite{jokisch_SPIE_2024}. One could envision using the platform, represented in \hyperref[fig:2]{Figure 5.b}, integrated with an optical detection scheme to trap and detect particles in situ in biological environments, opening a path for the development of new experiments and technologies in integrated biophotonics systems.

\section{Discussion}
The omnidirectional trapping of sub-wavelength nanoparticles in integrated optical devices is central to many applications in the field of optical trapping, such as microbiology \cite{sudhakar_germanium_2021, ashkin_optical_1987}, biophysics \cite{bustamante_optical_2021}, or fundamental physics \cite{gieseler_subkelvin_2012, dholakia_colloquium_2010, donato_light-induced_2016}. For deeply sub-wavelength particles, where the dipole approximation is valid, we demonstrate that it is possible to tailor the electric field distribution and the trapping potential, by careful nanostructuring of the dielectric environment. The target distribution of the electric fields results in a particle size agnostic omnidirectional optical trap based on gradient forces. We test the geometric limits of omnidirectional trapping by means of a parametric study that indicates omnidirectional trapping only for exclusion radii above $R_\text{exc} \geq 250$ nm  and demonstrate how CP forces also set a lower size limit for the exclusion radius ($R_\text{exc}$). To compare the optimized device to other nanophotonic platforms we propose the metric of normalized trapping stiffness ($\eta_i$), which shows unprecedented values compared to other state-of-the art optical tweezer platforms \cite{gieseler_subkelvin_2012}, highlighting the deeply sub-wavelength nature of the trapped particles and the low input power required for trapping. Moreover, our framework enables the design of manufacturable \cite{albrechtsen_nanometer-scale_2022} optical traps for a given application by rescaling the wavelength and the spatial dimensions of the device and adjusting the material parameters accordingly, before applying the inverse design process. We have demonstrated this in the short-wave infrared ($\lambda=1.55$ \textmu m) and near-infrared frequencies ($\lambda=775$ nm). Since the designs can be fabricated by standard electron-beam lithography, we anticipate experimental realizations of the optical traps, with applications in levitated cavity optomechanics and integrated biophotonics technologies.

We also foresee developments in our inverse design framework. For instance, modifications to the FOM and the optimization framework can enable the design of optical traps based on different materials, traps for lossy or resonant particles, traps with multiple trapping spots, or novel optical traps for multiple quantum emitters, such as quantum dots or cold atoms. These networks of trapped emitters may, in turn, be used to study and develop quantum many-body systems \cite{chang_colloquium_2018}.
\backmatter


\bmhead{Acknowledgements}

We would like to thank Jesper Mørk for the useful discussions and support. BFG, PTK, and MW would also like to thank Kishan Dholakia for the useful discussions. We gratefully acknowledge financial support from the Danish National Research Foundation through NanoPhoton - Center for Nanophotonics, grant number DNRF147.

\bibliography{TrappingBenatProj}



\newpage 

\newcommand*\mycommand[1]{\texttt{\emph{#1}}}
\renewcommand{\thefigure}{S\arabic{figure}} 
\renewcommand{\thetable}{S\arabic{table}} 
\renewcommand{\thesection}{S\arabic{section}} 
\renewcommand{\thesubsection}{S\arabic{section}.\arabic{subsection}} 
\renewcommand{\thepage}{S\arabic{page}}
\setcounter{equation}{0} 
\setcounter{figure}{0} 

\renewcommand{\theequation}{S\arabic{equation}} 
\def\figureautorefname{Figure}

\begin{center}
    {\Large Supplementary information for\par}
    \vspace{0.5cm}
    {\large Omnidirectional gradient force optical trapping in dielectric nanocavities by inverse design\par}
    \vspace{0.5cm}
\end{center}

\section*{S1 Forces in optical trapping}\label{sec:opt_trap}

To design a nanostructure that can omnidirectionally trap single particles we need to perform optical force calculations. For particles with a radius $R \ll \lambda$, where $\lambda$ is the wavelength of light, one may apply the dipole approximation. Calculating the cycle-averaged force acting on a point-dipole for monochromatic electromagnetic fields yields \cite{novotny_principles_2012}:
\begin{equation}
\langle\mathbf{F}\rangle = \mathbf{F}_\text{grad} + \mathbf{F}_\text{rad} + \mathbf{F}_{\text{SC}}\,,
\end{equation}
where $\mathbf{F}_\text{grad}$ is the gradient force, $\mathbf{F}_\text{rad}$ is the radiation pressure force and $\mathbf{F}_\text{SC}$ is the spin-curl force \cite{albaladejo_scattering_2009}. The gradient force is given by:
\begin{equation}
\mathbf{F}_\text{grad} = \frac{\alpha_{\text{R}}}{4} \nabla [\mathbf{E}^*(\mathbf{r})\cdot \mathbf{E}(\mathbf{r})]\,,
\end{equation}
where $\mathbf{E}(\mathbf{r})$ is the complex electric field, $\alpha_{\text{R}}$ and is the real part of the polarizability. The radiation pressure force is given by:
\begin{equation}
\mathbf{F}_\text{rad} = \frac{\sigma_\text{p}}{c}  \langle \mathbf{S} \rangle\,,
\end{equation}
where  $\sigma_\text{p} = \alpha_{\text{I}} \frac{k}{\varepsilon_0}$ is the particle's total cross-section, $\alpha_{\text{I}}$ is the imaginary part of the polarizability, $k$ is the wave-number, $\varepsilon_0$ is the vacuum permittivity, $c$ is the speed of light, $\langle \mathbf{S} \rangle = \frac{1}{2}\operatorname{Re}\left\{ \mathbf{E} \times \mathbf{H}^*\right\}$ is the cycle-averaged Poynting vector, and $\mathbf{H}$ is the complex magnetic field. The spin-curl force is given by:
\begin{equation}
\mathbf{F}_\text{SC} = \sigma_\text{p} c \left[ \nabla \times  \langle \mathbf{L} \rangle \right]\,,
\end{equation}
where  $\langle \mathbf{L} \rangle = \frac{\varepsilon_0}{4i\omega}\, \mathbf{E} \times \mathbf{E}^*$ is the cycle-averaged spin density of the electromagnetic field and $\omega$ is the angular frequency.

For spheres in a background medium with permittivity $\varepsilon_\text{back}$, the total polarizability is given by
\begin{equation}
    \alpha = \frac{\alpha_0}{1-\text{i}\frac{k^3}{6\pi\varepsilon_0}\alpha_0} = \alpha_{\text{R}} + \text{i} \alpha_{\text{I}}\,, \quad \text{with} \quad \alpha_0 = 4 \pi \varepsilon_0 R^3 \frac{\varepsilon-\varepsilon_\text{back}}{\varepsilon+2\varepsilon_\text{back}}\,,
\end{equation}
where $\alpha_0$ is the Clausius-Mossotti polarizability \cite{novotny_principles_2012} and  $\varepsilon(\omega)$ is a frequency-dependent dielectric permittivity. For lossless and non-resonant materials $\alpha_{\text{R}} \gg \alpha_{\text{I}}$ and the force can be completely described by the conservative gradient force, which is generated by a potential:
\begin{equation}\label{eq:trap_pot}
    U(\mathbf{r})=-\frac{\alpha_{\text{R}}}{4}[\mathbf{E}^*(\mathbf{r})\cdot \mathbf{E}(\mathbf{r})]\,,
\end{equation}
and the real part of the polarizability is described entirely by the Clausius-Mossotti equation $\alpha_\text{R} \simeq \alpha_0$. Note that to obtain an omnidirectional trapping potential, given that the force is $\mathbf{F}(\mathbf{r})=-\nabla U(\mathbf{r})$, it requires that the expression in \autoref{eq:trap_pot} has a single minimum in all spatial directions. The particle will experience a net zero force when it is at the minimum of the potential landscape, while if the particle is displaced from there in any direction, it will feel an attractive force pushing it back to the minimum, yielding omnidirectional trapping.

Calculating the cycle-averaged force acting on a particle directly from Maxwell's equations is also possible, without employing the dipole approximation. For a particle enclosed by the surface $\partial V$  the force is given by \cite{novotny_principles_2012}:
\begin{equation}\label{eq:MST_force}
\langle\mathbf{F}\rangle=\int_{\partial V}\langle\stackrel{\leftrightarrow}{\mathbf{T}}(\mathbf{r}, t)\rangle \cdot \mathbf{n}(\mathbf{r}) \mathrm{d} a
\end{equation}
where $\mathbf{n}$ defines the vector normal to the particle surface and $\stackrel{\leftrightarrow}{\mathbf{T}}$ is known as the Maxwell stress tensor and can be written as
\begin{equation}
\stackrel{\leftrightarrow}{\mathbf{T}} (\mathbf{r}, t) =\left[\varepsilon_0 \varepsilon \, \mathbf{\mathcal{E}}(\mathbf{r}, t) \otimes \mathbf{\mathcal{E}}(\mathbf{r}, t)+\mu_0 \mu \,\mathbf{\mathcal{H}}(\mathbf{r}, t) \otimes \mathbf{\mathcal{H}}(\mathbf{r}, t)-\frac{1}{2}\left(\varepsilon_0 \varepsilon \mathbf{\mathcal{E}}(\mathbf{r}, t)^2+\mu_0 \mu \mathbf{\mathcal{H}}(\mathbf{r}, t)^2\right) \stackrel{\leftrightarrow}{\mathbf{I}}\right]\,,
\end{equation}
where $\mathbf{\mathcal{E}}(\mathbf{r}, t)$  and $\mathbf{\mathcal{H}} (\mathbf{r}, t)$ are the time-dependent electric and magnetic fields respectively, $\otimes$ is the outer product, $\varepsilon$ and $\mu$ are the relative dielectric permittivity and permeability of the medium surrounding the particle, and $\varepsilon_0$ and $\mu_0$ are the free-space permittivity and permeability. The expression in \autoref{eq:MST_force} has been used to calculate the force acting on a particle introduced into the optimized optical trap to validate the results in the dipole approximation.

\section*{S2 Inverse design framework}\label{inverse}

\subsection*{S2.1 The forward problem}\label{sec:geom}

\begin{figure}[h!]
  \hspace*{-1.00cm} 
  \centering
  \includegraphics[scale=0.35]{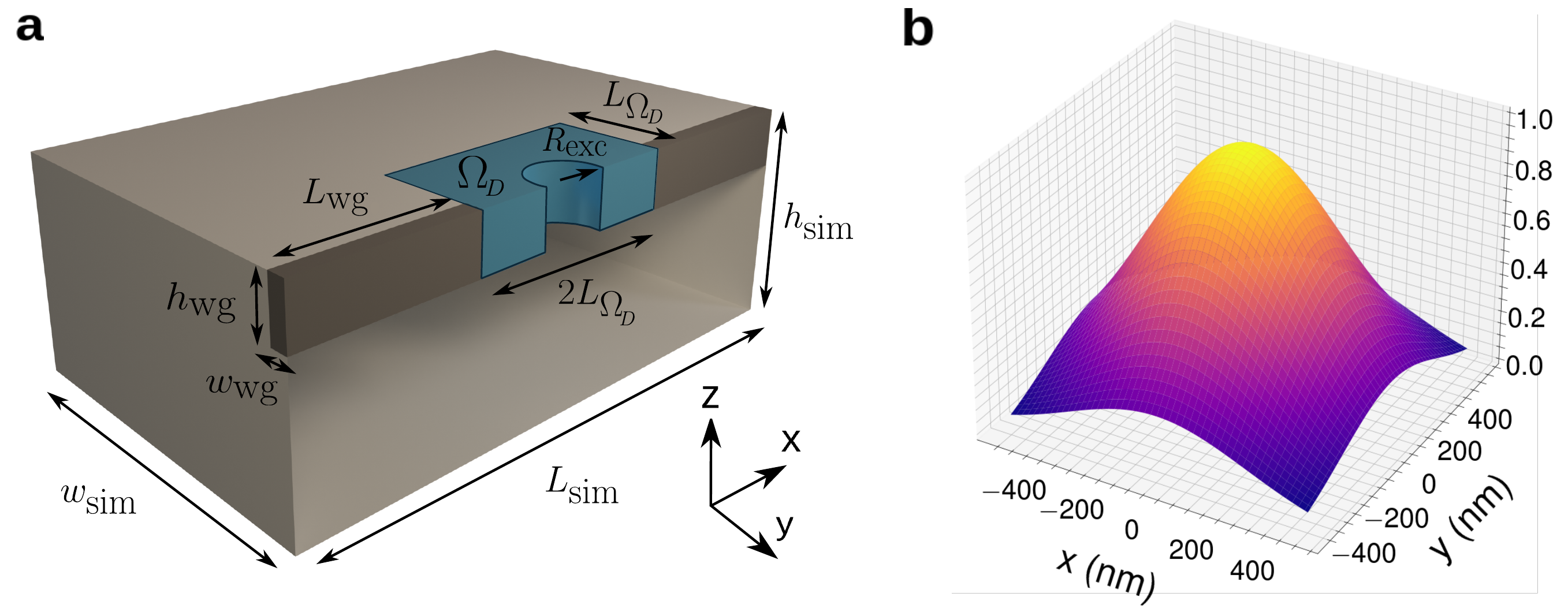}
  \caption{\textbf{a} Simulation domain with square design domain ($\Omega_D$) in blue connected to the optical waveguides. \textbf{b} Normalized reference Gaussian electric field for the $(x,y)$ plane.}
  \label{fig:S1}
\end{figure}

To inverse design the optical trap, we first need to define and solve an appropriate forward problem in order to model the nanophotonic system. This means solving Maxwell’s equations in the frequency domain, assuming time-harmonic field behavior \cite{novotny_principles_2012}. The model in \hyperref[fig:S1]{Figure S1.a} is discretized and solved using the finite element method with first-order Nedelec elements \cite{jin_finite_2014}. The model is based on a simulation domain consisting of two silicon optical waveguides (in dark gray) connected to a design region $\Omega_D$ (in blue) and an air cladding (in light gray). For the design in the near-infrared ($\lambda=1.55$\textmu m), the dimensions ($x$, $y$, $z$) for the simulation domain are $(L_\text{sim}, w_\text{sim}, h_\text{sim})=(9.1\, $\textmu m$, 4 \,$\textmu m$, 1.5\,$\textmu m$)$, which are large enough to allow the fields to decay away from the cavity, avoiding artificial numerical boundary reflections\footnote{This is also the case at the waveguide ends. The waveguide is excited with the fundamental mode at a port located a wavelength ($\lambda$) away from the edge of the simulation domain. This allows the fields reflected in the waveguide to decay away from the input port.}. The waveguide has dimensions of $(L_\text{wg}, w_\text{wg}, h_\text{wg})=(2.8 \, $\textmu m$, 275 \,\text{nm}, 400\, \text{nm})$. The height of the waveguide was chosen to be at least half of the wavelength since this was the smallest value yielding omnidirectional trapping. The dimensions for the design domain are $(2L_{\Omega_D},L_{\Omega_D}, h_\text{wg})=(2.5 \, $\textmu m$, 1.25 \, $\textmu m$, 400\, \text{nm})$ and are chosen to have a footprint of around $5(\lambda/n)^2$, where $n$ is the refractive index of silicon. This device size is large enough for the topology optimization to design couplers and mirrors. Lastly, in the center of the design domain, there is a cylindrical exclusion region to trap the particle, with radius $R_\text{exc}=300$ nm. This radius has been carefully chosen to obtain omnidirectional optical traps since there is a trade-off between the compactness of the exclusion region radius and the possibility of obtaining omnidirectional trapping, as further discussed in the main text. Note that for the integrated biophotonic design the wavelength and all dimensions are halved.

To solve the problem in a computationally efficient way, we assume that the device is symmetric around the $(x,y)$ and $(x,z)$ planes, leaving only a quarter of the total simulation domain to be solved. To impose the symmetry we apply perfect electric conductor (PEC) boundary conditions on the $(x,z)$ plane and perfect magnetic conductor boundary (PMC) conditions on the $(x,y)$ plane. On the rest of the boundaries, we apply first-order absorbing boundary conditions. Lastly, PML regions with a length of $\lambda$ are defined at the ends of both waveguides to avoid reflections at the ends of the simulation domain. 

\subsection*{S2.2 The optimization problem}\label{sec:opt_prob}

Having computed the solution to the forward problem for a given material distribution in the modeling domain, we compute the Figure of Merit (FOM) $\Phi$, which is to be optimized. As outlined in the main document, the FOM defines the difference of the electric-field norm with respect to a reference field, which is directly related to the shape of the potential and is given by:
\begin{equation}\label{eq:FOM_S}
\text{FOM} \equiv \Phi= \sqrt{\int_{\Omega} \left[ \Theta\left(\frac{\left\|\mathbf{E}(\mathbf{r})\right\|}{\left\|\mathbf{E}(\mathbf{r}_0)\right\|}-\frac{\left\|\mathbf{E}_\text{ref}(\mathbf{r})\right\|}{\left\|\mathbf{E}_\text{ref}(\mathbf{r}_0)\right\|}\right)\right]^2 \mathrm{~d} \Omega}\,,
\end{equation}
where $\mathbf{r}_0=(x_0, y_0, z_0) = (0,0,0)$ is the center of the design domain, $\mathbf{E}_{\text{ref}}$ is a reference electric field, $\Omega$ is the optimization domain defined by the exclusion region in \hyperref[fig:S1]{Figure S1.a}, and $\Theta(x)$ is a smoothed Heaviside threshold function \cite{wang_projection_2011}. The target reference field is chosen to be a three-dimensional Gaussian of the form:
\begin{equation}\label{eq:gaussian}
    \frac{\left\|\mathbf{E}_\text{ref}(\mathbf{r})\right\|}{\left\|\mathbf{E}_\text{ref}(\mathbf{r}_0)\right\|} = \exp \left(-\left(\frac{\left(x-x_0\right)^2}{2 \sigma_x^2}+\frac{\left(y-y_0\right)^2}{2 \sigma_y^2} + \frac{\left(z-z_0\right)^2}{2 \sigma_z^2}\right)\right)\,.
\end{equation}
where $\sigma_{x} = \sigma_{y} = 300$ nm and $\sigma_{z}= 400$ nm are the standard deviations in all directions, chosen to match the exclusion size. In \hyperref[fig:S1]{Figure S1.b} we show a projection of the field described by \autoref{eq:gaussian} for the $(x,y)$ plane, which showcases that the Gaussian has a single maximum at $\mathbf{r}=\mathbf{r}_0$. As pointed out in \hyperref[sec:opt_trap]{Section S1}, if the electric-field norm has a single maximum, this will result in a global minimum of the trapping potential, which is necessary to trap single particles omnidirectionally. 

The FOM is minimized by optimizing the material distribution of silicon (Si) and air in the design region ($\Omega_D$). We formulate the design problem as a continuous optimization problem, where the material distribution is controlled by a design field $\xi$, which is discretized into a piecewise constant field coinciding with the finite elements used to discretize the physics model, with one design variable controlling the value of the design field in each element. To regularize the design and enable length-scale control, we adopt a filtering and
thresholding scheme. For the filter, we use a Helmholtz-based filter \cite{lazarov_filters_2011}:
\begin{equation}\label{eq:filter}
-\left(\frac{r_f} {2\sqrt{3}}\right)^2\nabla\tilde{\xi}+\tilde{\xi}=\xi\,,
\end{equation}
where $\tilde{\xi}$ is the filtered design field, and $r_f$ is the filter radius. The filter operation is followed by a smoothed Heaviside threshold ($\Theta$) \cite{wang_projection_2011}:
\begin{equation}\label{eq:thres}
    \bar{\tilde{\xi}} = \Theta(\tilde{\xi}
    ) = \frac{\tanh (\beta \cdot \eta)+\tanh (\beta \cdot(\tilde{\xi}-\eta))}{ \tanh (\beta \cdot \eta)+\tanh (\beta \cdot(1-\eta))}, \quad \beta \in[1, \infty), \eta \in[0,1]\,,
\end{equation}
where $\bar{\tilde{\xi}}$ is the filtered and thresholded design field, and $\beta$ and $\eta$ control the threshold sharpness and value respectively. To translate the design field into the material distribution in the physics model we employ a non-linear material
interpolation \cite{christiansen_non-linear_2019, jokisch_topology_2024}:
\begin{equation}
\begin{aligned}
& \varepsilon_r(\overline{\tilde{\xi}})=\left(n(\overline{\tilde{\xi}})^2- K(\overline{\tilde{\xi}})^2\right)-\mathrm{i}(2 n(\overline{\tilde{\xi}}) K(\overline{\tilde{\xi}}))-\mathrm{i} a \overline{\tilde{\xi}}(1-\overline{\tilde{\xi}}), \\
& n(\overline{\tilde{\xi}})=n_{\text {air }}+\overline{\tilde{\xi}}\left(n_{\text {Si }}-n_{\text {air }}\right), \\
& K(\overline{\tilde{\xi}})=K_{\text {air }}+\overline{\tilde{\xi}}\left(K_{\text {Si }}-K_{\text {air }}\right),
\end{aligned}
\end{equation}
where $\varepsilon_r$ is the relative dielectric permittivity, $n$ is the refractive index, $K$ is the extinction coefficient, $a$ is a problem-dependent parameter known as the artificial attenuation, and  ``Si" stands for silicon. With this setup we optimize the design using the method of moving asymptotes \cite{svanberg_method_1987}  with a single homogeneous design guess ($\xi = \xi_0$), leaving the optimizer free to tailor the device geometry. The optimizer solves the following optimization problem:
\begin{subequations}
\begin{align}
 \min _{\xi}: \,\, & \text{FOM} \equiv \Phi= \sqrt{\int_{\Omega} \left[ \Theta\left(\frac{\left\|\mathbf{E}(\mathbf{r})\right\|}{\left\|\mathbf{E}(\mathbf{r}_0)\right\|}-\frac{\left\|\mathbf{E}_\text{ref}(\mathbf{r})\right\|}{\left\|\mathbf{E}_\text{ref}(\mathbf{r}_0)\right\|}\right)\right]^2 \mathrm{~d} \Omega}\,, \label{eq:optimization:FOM}\\ \text { s.t. : } &\mathbf{S}\left(\varepsilon_r(\bar{\tilde{\xi}}, \mathbf{r})\right) \mathbf{E}(\mathbf{r}) = \mathbf{F}(\mathbf{r}) \,, \label{eq:optimization:maxwell}\\
  :\,\, & \log_{10}\left({||\mathbf{E}(\mathbf{r}_0)||}\right)\geq \gamma \,,\label{eq:optimization:intensity}\\
  :\,\, & \nabla \cdot (-\sigma_C \,\, \Theta\left(\bar{\tilde{\xi}}(\mathbf{r})\right) \nabla C(\mathbf{r}) ) =f\,\Theta \left(\bar{\tilde{\xi}}(\mathbf{r})\right), \quad C=0 \,\,\, \forall \mathbf{r} \in \Gamma_i, \quad i\in(1,2)\,, \label{eq:optimization:connec}\\
:\,\, & M_C=\int_{\Omega_D} C \mathrm{~d} \mathbf{r} \leq \epsilon_C \label{eq:connec_con}\\
  : \, \, & g^s = \frac{1}{n} \sum_{i \in \mathbb{N}} I^s_i [\text{min}\{(\tilde{\xi}_i - \eta_e), 0\}]^2\, \leq \epsilon\,, \label{eq:optimization:l1} \\
  : \, \, & g^v = \frac{1}{n} \sum_{i \in \mathbb{N}} I^v_i [\text{min}\{(\eta_d - \tilde{\xi}_i ), 0\}]^2\, \leq \epsilon \,, \label{eq:optimization:l2}\\
 :\,\, & 0<\xi(\mathbf{r})<1 \,,\\
 :\,\, & \xi=0 \quad \forall \mathbf{r} \in \Omega_D \, ,
\end{align}
\end{subequations}
where the FOM in \autoref{eq:optimization:FOM} is subject to the constraints given by the individual subequations:
\begin{itemize}
    \item \autoref{eq:optimization:maxwell} is the discretized form of Maxwell's equations, which is equivalent to solving a linear algebraic system that yields the electric field for the system matrix $\mathbf{S}$ and the excitation term $\mathbf{F}$.
    \item \autoref{eq:optimization:intensity} is a constraint for the electric-field norm in the center of the domain, where $\gamma$ is a problem-dependent parameter. By selecting $\gamma=1.15 \log_{10}\left(||\mathbf{E}(\mathbf{r}_0)||_{i=0}\right)$, where $i=0$ refers to the initial design, we ensure that in the initial steps of the optimization, the optimizer avoids local minima where the trapping potential has the correct shape but is not deep enough for stable trapping. Once the constraint in \autoref{eq:optimization:intensity} is fulfilled, it ensures a stable trapping potential that overcomes thermal diffusion, and the optimizer directly targets the FOM in \autoref{eq:optimization:FOM} to achieve the correct shape of the trapping potential.
    \item \autoref{eq:optimization:connec} refers to two connectivity equations formulated using an artificial heat-transfer problem, that ensures a design connected to the two waveguides. In this expression, $\sigma_C$ is a material interpolation for the artificial conductivity, $C$ denotes the artificial temperature field, $f$ denotes the artificial heat generated by materials and  $\Gamma_i$ describes the heat sink boundaries given by our two ($i=1,2$) waveguide ends \cite{li_structural_2016}. The rest of the parameters are selected as detailed {in \cite{christiansen_inverse_2023}}.
     \item \autoref{eq:connec_con} refers to the connectivity constraint $M_C$  where $\epsilon_C$ is a sufficiently small to ensure that all material is connected to the boundaries, as detailed {in \cite{christiansen_inverse_2023}}.
    \item  The constraints $g_s$ and $g_v$, in \autoref{eq:optimization:l1} and \autoref{eq:optimization:l2}, are the solid and void connectivity constraints, where $\eta_e$ and $\eta_d$ are the eroded and dilated thresholds, $n$ denotes the number of elements, $I^s_i=\bar{\tilde{{\xi}}}_i \cdot \text{e}^{-p\cdot|\nabla\tilde{\xi}_i|^2}$ and $I^v_i=(1-\bar{\tilde{{\xi}}}_i) \cdot \text{e}^{-p\cdot|\nabla\tilde{\xi}_i|^2}$ are the solid and void indicator functions, $p$ is a problem-dependent parameter and $\epsilon$ is the length scale error \cite{zhou_minimum_2015}. To ensure the minimum length scale of the design features we select $p = r_f^4$ \cite{zhou_minimum_2015}, where $r_f$ is the filter radius, and $\eta_e=0.75$, $\eta_d=0.25$ which in our optimization problem ensure that no features in the design have a radius of curvature smaller than 60 nm \cite{hammond_photonic_2021}, making the design manufacturable using e.g. standard electron beam lithography.
\end{itemize}

We solve the optimization problem in \hyperref[eq:optimization:FOM]{Equations S14a-S14h} starting from a single homogenous initial guess of the design field $\xi=0.6$ for all the design variables. Additionally, we use a continuation scheme to exploit the continuous nature of the design field in solving the optimization problem while achieving a final, physically realizable binarized design. This consists of increasing the parameter that controls the threshold sharpness ($\beta$) and the artificial attenuation ($a$) every 50 iterations of the optimization, pushing the design field toward binary values and thus physical realizability. We also gradually reduce the length scale error ($\epsilon$) introduced in \hyperref[eq:optimization:FOM]{Equations S14e-S14f} to ensure that the final design fulfills the minimum length scale requirement. The continuation scheme parameters are summarized in \autoref{tabcont}. 

\begin{table}[htbp]
\caption{Continuation scheme parameters in the topology optimization framework.}\label{tabcont}%
\footnotesize
\begin{tabular}{@{}l|llllllllll@{}}
\toprule
	 \textbf{Continuation step} & $0$ & 1 & 2  & 3 & 4 & 5 & 6 & 7 & 8 & 9   \\
  \midrule
  Iteration & $0$ & 50 & 100  & 150 & 200 & 250 & 300 & 350 & 400 & 450   \\
  Threshold sharpness (\textbf{$\beta$}) & $5$ & 7.5 & 10  & 15 & 25 & 35 & 50 & 75 & 100 & 150   \\
  Artificial attenuation (\textbf{$a$}) & $0.01$ & 0.1 & 0.2  & 0.4 & 0.8 & 0.8 & 0.8 & 0.8 & 0.8 & 0.8   \\
  Length scale error ($\epsilon$) & 1 & 1 & 1  & 1 & $10^{-3}$ & $7.5 \cdot 10^{-4}$ & $5 \cdot 10^{-4}$ & $5 \cdot 10^{-4}$ & $5 \cdot 10^{-4}$ & $5 \cdot 10^{-4}$
	\end{tabular}
\end{table}

\section*{S3 Surface forces}\label{forces}
A particle located at the center of our optimized device will be omnidirectionally trapped. There are, however, two physical effects that could potentially compromise loading the particle into the optical trap: optical lightning-rod effects \cite{albrechtsen_two_2022, choi_self-similar_2017} and vacuum fluctuation-induced Casimir-Polder (CP) forces \cite{casimir_influence_1948}. The CP forces are studied in the main text, while in this section we focus on the lightning-rod effects.

\subsection*{S3.1 Lightning-rod effects}
By taking a close look at the field distribution at the center of the nanocavity in Figure 1.a in the main document, one can see a significant local field enhancement at the material interface for $z = -400$ nm. This is an inherent consequence of the boundary conditions of Maxwell's equations \cite{choi_self-similar_2017}, where the field is locally enhanced
at kinks and corners \cite{albrechtsen_two_2022}. This so-called lightning-rod effect is strongly confined to the interface but still creates an attractive force on the particle if the particle is sufficiently far from the center, and sufficiently close to the interface. 

\begin{figure}[h!]
  \hspace*{-1.00cm} 
  \centering
  \includegraphics[scale=0.35]{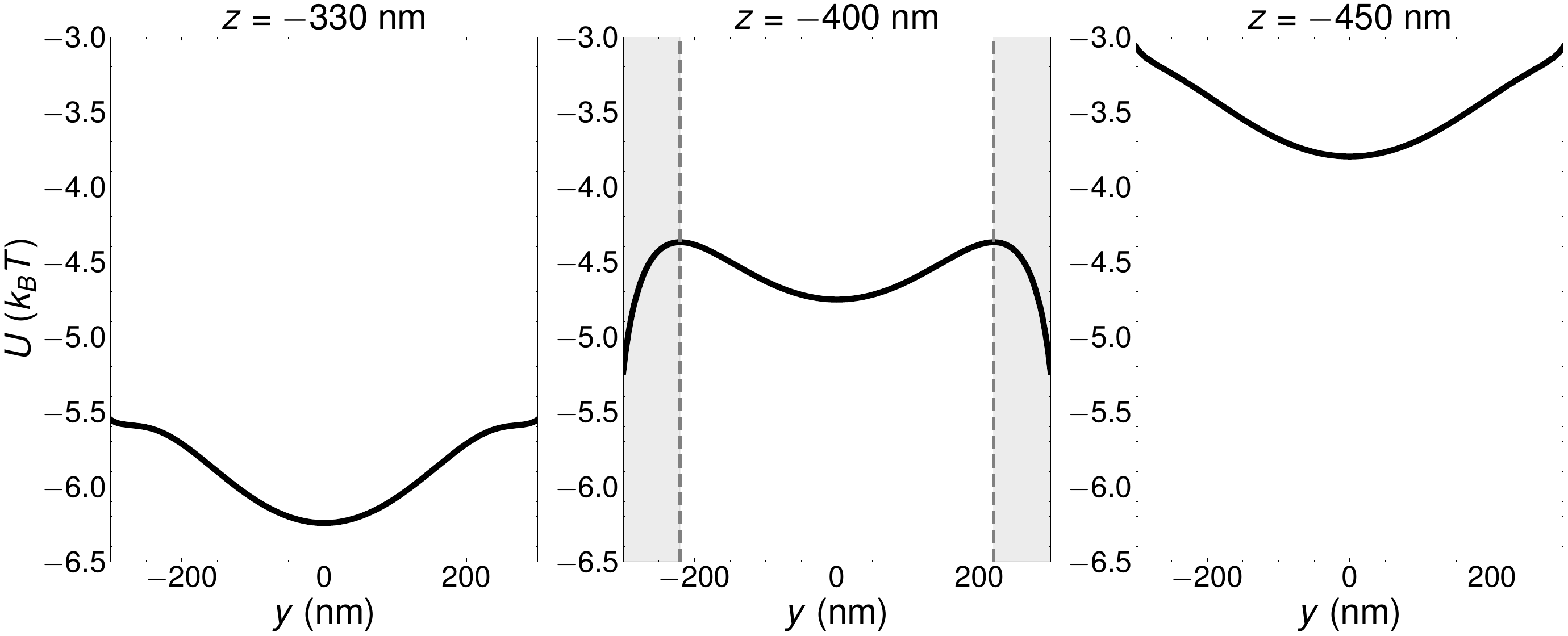}
  \caption{Trapping potential for the $y$ line in the optimization region at different out-of-plane heights ($z$). The region where the particle could stick to the interface is marked in gray.}
  \label{fig:2_S}
\end{figure}

To quantify this effect, we investigate the trapping potentials for the $y$ line at different out-of-plane ($z$ axis) heights in \autoref{fig:2_S}\footnote{These results have been validated for a 15 nm radius particle using the MST formalism.}. At the bottom of the device ($z=-400$ nm), we see that there is a region $y \in[-220\text{ nm}, 220\text{ nm}]$ where the particle will be stably trapped, feeling a restorative force towards the center. In the other region (in gray) the particle will feel an attractive force towards the interface, which will lead to the particle sticking to the interface. If we go up inside the structure ($z=-330$ nm), or down out of the structure ($z=-450$ nm) we see the sticking effect disappearing. This means that if we assume an initial random position of the particle above the hole in the simulation domain in \autoref{fig:S1}, we can compute the probability that it will get stuck at the interface as it travels into the omnidirectional trap. We do this by assuming a simple positional averaging for the particle as it travels into the trap and then calculating the volume ratio between the volume where the particle feels an attractive force to the interface and the total volume. As we go up or down the structure, the region where the particle will get stuck to the interface (in gray in \autoref{fig:2_S}), becomes smaller until it disappears entirely as we reach the limits $z=-450$ nm and $z=-330$ nm. To simplify the calculations and provide a worst-case estimate, we assume that the width of this region in the $y$ axis remains constant as we move up and down in the structure. With this assumption the probability of sticking is
\begin{equation}
    P_\text{stick}^\text{LR} \, [\%]= \frac{V_\text{stick}}{V_\text{total}} = \frac{ h_\text{stick}(R^2_\text{exc} - R^2_\text{stick} )}{ h_\text{sim} R^2_\text{exc}} \simeq 4.5 \% \,,
\end{equation}
where ``LR" denotes lightning-rod, $h_\text{stick}=120$ nm is the height of the region where the particle can stick to the surface, in the interval  $z\in[-450\text{ nm}, -330\text{ nm}]$. Therefore, there is a small probability that the particle might stick to the interface due to the lightning-rod effect at the top and bottom interfaces of the structure, as it is being loaded into the trap. It is noted that this is a conservative approximation since it assumes that the unstable effect does not fade out as we go into or out of the structure and does not account for the inherent velocity of the particle, or other forces that could pull the particle out of the unstable region. Thus, our estimate constitutes a worst-case sticking probability $P_\text{stick}^\text{LR}$, and a significantly smaller value is expected in reality.  

\end{document}